\documentclass[twocolumn,prl,aps,showpacs]{revtex4}
\usepackage{graphicx}

\begin{document}

\title{The Ehrenfest Urn Model with Interaction  }

\author{Chun-Hsiung Tseng}
\author{Yee-Mou Kao}
\author{Chi-Ho Cheng}
\email{phcch@cc.ncue.edu.tw}

\affiliation{Department of Physics, National Changhua University
of Education, Taiwan}

\date{\today}

\begin{abstract}
We studied the Ehrenfest urn model in which particles in the same
urn interact with each other. Depending on the nature of
interaction, the system undergoes a first-order or second-order
phase transition. The relaxation time to the equilibrium state,
the Poincare cycles of the equilibrium state and the most
far-from-equilibrium state, and the duration time of the states
during first-order phase transition are calculated. It was shown
that the scaling behavior of the Poincare cycles could be served
as an indication to the nature of phase transition, and the ratio
of duration time of the states could be a strong evidence of the
metastability during first-order phase transition.
\end{abstract}


\maketitle


\section{I. Introduction}

Historically, the Boltzmann's $H$ theorem based on the assumption
of molecular chaos singles out a direction of time, which led to
two pardoxes \cite{huang}. The first one, so-called reversal
paradox, states that the $H$ theorem is inconsistent with the time
reversal invariance. The Poincare theorem \cite{poincare} requires
that the system should return to its initial state (up to an
arbitrarily small neighborhood) after sufficiently long time. This
fact implies reversibility of the dynamical system, leading to the
so-called recurrence paradox. Later on, the Ehrenfest urn model
\cite{ehrenfest} was proposed to resolve the paradoxes and clarify
the relationship between reversible dynamics and irreversible
thermodynamics.

The Ehrenfest model deals with two urns with total $N$ particles.
Each particle is randomly chosen with equal probability in such a
way that it is taken from one urn to another urn. It is found that
the relaxation time for the system to reach its equilibrium is
proportional to $N$, and the Poincare cycle of the most
far-from-equilibrium state is proportional to $2^N$ \cite{kac}.

Since then, the Ehrenfest model was generalized such that the
jumping rates between two urns are unbalanced
\cite{siegert,klein}, the system of two urns becomes multiurn
\cite{kao03,kao04,nagler05}, and multiurn are connected in a
complex network \cite{clark}. Fluctuation distribution of the
model was also studied \cite{hauert,bakar09,bakar10}.

The Ehrenfest model was also applied to understand the granular
system by inducing different effective temperatures with respect
to gravitational field in different urns, which turns out to
exhibit the spatial separation (symmetry-breaking) phase
transition \cite{lipowski02a,lipowski02b,coppex}. This model was
also solved analytically \cite{shim}.

By considering the continuum limit of time step in the evolution
of the probability of the state, the linear Fokker-Planck equation
is obtained \cite{kac,nagler99}. Modification of the Ehrenfest
model by incorporating nonlinear contribution to the Fokker-Planck
equation recently calls for attention \cite{curado,nobre,casas},
which is motivated by the processes associated with
anomalous-diffusion phenomena \cite{bouchaud,boon,lutsko}. The
generalized $H$ theorem for the nonlinear Fokker-Planck equation
was studied by many authors in recent years
\cite{schwammle,shiino,frank,chavanis}.

Although many attempts were made to modify the Ehrenfest model,
none of them has been associated with {\it explicit} particle
interaction, to our knowledge. The model that we modified exhibits
the (first-order and second-order) phase transition depending on
the nature of interaction. We also calculate the relaxation time
to the equilibrium state, the Poincare cycles of both the
equilibrium and the most far-from-equilibrium states, and the
duration time of the states during the first-order phase
transition. Finally, we point out that the scaling behavior of the
Poincare cycle could be served as an indication of the nature of
the phase transition, and the ratio of duration time of the states
could be a strong evidence of the metastability during first-order
phase transition.


\section{II. Ehrenfest model with interaction}

We present our model as follows. There are $N$ particles
distributed into two urns. The number of particle in the left and
right urns are $n$ and $N-n$, respectively. Since the total
particle number $N$ is fixed, we label the state of the system by
its particle number in the left urn, denoted by $|n\rangle$.

Unlike the original Ehrenfest model, we introduce particle
interaction in the same urn. Two particles of different urns do
not interact. The total energy $E = \frac{J}{2}( n(n-1) +
(N-n)(N-n-1) )$ with energy coupling $J$. The interaction is
attractive (replusive) if $J$ is negative (positive). When a
particle jumps from the left to the right urn, $\Delta E =
-J(2n-N-1)$. To satisfy the principle of detailed balance, we
should have the restriction on the transition probability such
that
\begin{eqnarray}
\frac{T_{n,n-1}}{T_{n-1,n}} = {\rm e}^{\beta \Delta E} = {\rm
e}^{-\frac{g}{N}(2n-N-1)} \label{Tnn}
\end{eqnarray}
where $T_{n\pm1,n}$ is the transition probability from the state
$|n\rangle$ to $|n\pm1\rangle$, $\beta$ is the inverse of
effective temperature, and we introduce the coupling constant
$g\equiv N J \beta$ such that $\Delta E$ is extensive
(proportional to $N$ given fixed $g$). There is a degree of
freedom to choose the transition probability; however, we adopt
\begin{eqnarray}
T_{n-1,n} &=& \frac{1}{{\rm e}^{-\frac{g}{N}(2n-N-1)} + 1} \\
T_{n,n-1} &=& \frac{1}{{\rm e}^{\frac{g}{N}(2n-N-1)} + 1}
\end{eqnarray}
Note that $T_{n-1,n}=T_{n,n-1}=\frac{1}{2}$ if the interaction is
turned off. Different proportionality implies different time scale
chosen. Besides the particle interaction, we further introduce the
jumping rate from one urn to another urn, which is independent of
the particle interaction. Suppose the probability of jumping rate
from the left (right) to the right (left) urn is $p ( q )$. For
convenience, we restrict $p + q = 1$. Again this restriction only
changes the time scale.

\begin{figure}[tbh]
\begin{center}
\includegraphics[width=3in]{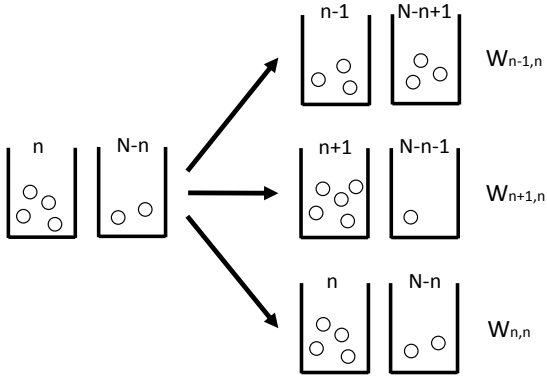}
\end{center}
\vspace{-5pt}
 \caption{Schematic Diagram to illustrate the transition in our model.}
 \label{schematic.eps}
 \vspace{25pt}
\end{figure}

After $s$ steps from the initial state $|n_0\rangle$, the
probability of the state $|n\rangle$ is denoted by $\langle n|
p(s) | n_0 \rangle$, where $p(s)$ is the corresponding operator.
As illustrated in Fig. \ref{schematic.eps}, one have the
recurrence relation from the $(s-1)$-th to $s$-th step such that
\begin{widetext}
\begin{eqnarray} \label{recurrence}
\langle n| p(s) | n_0 \rangle = W_{n,n-1} \langle n-1 | p(s-1) |
n_0 \rangle + W_{n,n+1} \langle n+1 | p(s-1) | n_0 \rangle  +
W_{n,n} \langle n | p(s-1) | n_0 \rangle
\end{eqnarray}
\end{widetext}
where $W_{n-1,n} = \frac{n}{N} p T_{n-1,n}$, $W_{n,n-1} =
\frac{N-n+1}{N} q T_{n,n-1}$, and $W_{n,n} = 1 - W_{n-1,n} -
W_{n+1,n}$.

\vspace{25pt}

It is convenient to rewrite the recurrence relation in a matrix
form. Let the state vector
\begin{eqnarray}
\psi(s) = \left(\begin{array}{c}
  \langle 0 |p(s)| n_0 \rangle \\
  \langle 1 |p(s)| n_0 \rangle \\
  \vdots \\
  \langle N |p(s)| n_0 \rangle \\
\end{array} \right)
\end{eqnarray}
The normalization condition (probability conservation) of the
state vector is $\sum_{n=0}^N \psi_n(s) = \sum_{n=0}^N \langle n
|p(s)|n_0 \rangle = 1$ for any $s$. Define the matrix $M_{nm} =
\langle n |p(1)| m \rangle$ so that $\psi(s) = M \psi(s-1)$. In
general, $\psi(s) = M^s \psi(0)$. Based on the normalization
condition of the state vectors, the matrix $M$ should satisfy
\begin{eqnarray}  \label{mnorm}
\sum_{n=0}^N (M^s)_{nm} = 1
\end{eqnarray}
for $m=0,1,\ldots,N$ and $s\geq 1$. $M^s$ can be evaluated if the
eigenvalues $\lambda_m$ and eigenvectors $\phi(m)=(\phi_0(m),
\phi_1(m), \ldots, \phi_N(m))^t$ of $M$ are known, and so
\begin{eqnarray}
M^s = A \Lambda^s A^{-1}
\end{eqnarray}
where $A$ and $\Lambda$ are matrices of dimension
$(N+1)\times(N+1)$. Their components are $A_{nm} = \phi_n(m)$ and
$\Lambda_{nm}=\lambda_m \delta_{nm}$.

The eigensystem becomes
\begin{widetext}
\begin{eqnarray} \label{eigensystem}
&& \frac{N-n+1}{N}\frac{q}{{\rm e}^{\frac{g}{N}(2n-N-1)}+1}
\phi_{n-1} + \frac{n+1}{N}\frac{p}{{\rm
e}^{-\frac{g}{N}(2n-N+1)}+1} \phi_{n+1} \nonumber \\ && + \left( 1
- \frac{n}{N}\frac{p}{{\rm e}^{-\frac{g}{N}(2n-N-1)}+1}
-\frac{N-n}{N}\frac{q}{{\rm e}^{\frac{g}{N}(2n-N+1)}+1} \right)
\phi_n = \lambda \phi_n
\end{eqnarray}
\end{widetext}
The indices $m$ to $\lambda$ and $\phi_n$ is omitted without
causing any confusion. We found no exact solution to the
eigenproblem except for some special cases, {\it e.g.} the cases
in which $g=0$ and $g\rightarrow -\infty$ (See Appendix A and B
for details). If $\lambda_N = 1$ (we label its index $N$),
\begin{eqnarray} \label{phinN}
\phi_n(N) = \frac{N!}{n! (N-n)!} p^{N-n} q^n {\rm
e}^{\frac{g}{N}n(N-n)}
\end{eqnarray}
in which the eigenstate could be verified by directly substitution
into Eq. (\ref{eigensystem}).

\section{III. Mean Particle Number}

The mean particle number after $s$ steps,
\begin{eqnarray} \label{ns}
\langle n \rangle_s &=& \sum_{n=0}^N n \psi_n(s) \nonumber \\
&=& \sum_{n=0}^N n (M^s \psi(0))_n \nonumber \\
&=& \sum_{n=0}^N \sum_{m,k=0}^N n A_{nm} \lambda_m^s A_{mk}^{-1}
\psi_k(0)
\end{eqnarray}
Suppose there is an unique state of unity eigenvalue, says,
$\lambda_N=1$, and all the remaining eigenvalues are less than
unity, as $s\rightarrow\infty$, the mean value $\langle n \rangle$
is defined as
\begin{eqnarray} \label{ninfty}
\langle n \rangle \equiv \langle n \rangle_\infty = \sum_{n=0}^N
\sum_{k=0}^N n A_{nN} A_{Nk}^{-1} \psi_k(0)
\end{eqnarray}
By taking the limit $s\rightarrow\infty$ in Eq. (\ref{mnorm}), we
get $\sum_{n=0}^N A_{nN} A_{Nm}^{-1} = 1$ for any $m$. Hence
\begin{eqnarray} \label{ainm}
A_{Nm}^{-1} = \frac{1}{\sum_{n=0}^N A_{nN}} =
\frac{1}{\sum_{n=0}^N \phi_n(N)}
\end{eqnarray}
which is independent of $m$. Substitute Eq. (\ref{ainm}) into Eq.
(\ref{ninfty}),
\begin{eqnarray} \label{nmean}
\langle n \rangle &=& \sum_{n=0}^N  n \phi_n(N)
\frac{1}{\sum_{n=0}^N \phi_n(N)} \sum_{k=0}^N \psi_k(0)  \nonumber
\\
&=& \frac{\sum_{n=0}^N  n \phi_n(N)}{\sum_{n=0}^N \phi_n(N)}
\end{eqnarray}

\begin{figure}[tbh]
\begin{center}
\includegraphics[width=3in]{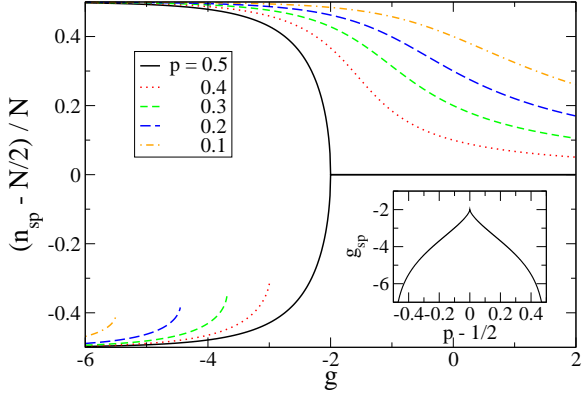}
\end{center}
\vspace{-5pt}
 \caption{The relation between the saddle
point $y_{\rm sp}$ in Eq. (\ref{y}) and the coupling constant $g$.
$\frac{1}{N}(n_{\rm sp} -\frac{N}{2})$ as a function of coupling
constant $g$ for different $p$.
 As $g \leq g_{\rm sp}$, two saddle points arise where $n_{\rm sp,-} < n_{\rm sp,+}$.
 Inset: $g_{\rm sp}$ as a function of
 $p-\frac{1}{2}$.
  }
 \label{y_vs_g.eps}
 \vspace{25pt}
\end{figure}

In general, there is no closed form for Eq. (\ref{nmean}) if $N$
is finite. If $N$ is large enough, we could derive the asymptotic
result. Notice that, by using the Stirling formula \cite{mathews},
one can rewrite Eq. (\ref{phinN}) as $\phi_n(N) = \exp(N
f(\frac{n}{N})-\frac{1}{2}\log(2\pi
\frac{n}{N}(1-\frac{n}{N})N)+O(N^{-1}))$. Then the denominator in
Eq. (\ref{nmean})
\begin{eqnarray} \label{sum}
\sum_{n=0}^N \phi_n(N) = \left(\frac{N}{2\pi}\right)^{\frac{1}{2}}
\int_0^1 dx \frac{{\rm e}^{N f(x)}}{\sqrt{x(1-x)}}
\end{eqnarray}
where $x=\frac{n}{N}$, the proportion of particle number in the
left urn, and
\begin{eqnarray}
f(x) &=& -x\ln x - (1-x)\ln(1-x) \nonumber \\ && + (1-x)\ln p +
x\ln q + g x(1-x)
\end{eqnarray}
As $N$ is large enough, the integral is asymptotically
\begin{eqnarray}  \label{sumlargeN}
&& \left(\frac{N}{2\pi}\right)^{\frac{1}{2}} \sum_{\{ x_{\rm sp}
\} } \frac{{\rm e}^{N f(x_{\rm sp})}}{\sqrt{x_{\rm sp}(1-x_{\rm
sp})}} \int_0^1 dx {\rm e}^{\frac{N}{2}
f''(x_{\rm sp})(x-x_{\rm sp})^2} \nonumber \\
&=&  \sum_{\{ x_{\rm sp} \} } \frac{{\rm e}^{N f(x_{\rm
sp})}}{\sqrt{x_{\rm sp}(1-x_{\rm sp})} |f''(x_{\rm
sp})|^\frac{1}{2}}
\end{eqnarray}
where $\{ x_{\rm sp} \}$ is the set of the saddle points
satisfying $f'(x_{\rm sp})=0$ and $f''(x_{\rm sp}) < 0$. $x_{\rm
sp}$ represents the proportion of particle number in the left urn
at equilibrium state or metastable state. The condition that
$f'(x_{\rm sp}) = 0$ is expressed as
\begin{eqnarray} \label{y}
2 y_{\rm sp} = -\tanh\left[ g y_{\rm sp} + \frac{1}{2}\ln
\left(\frac{p}{q}\right)\right]
\end{eqnarray}
where $y_{\rm sp} \equiv x_{\rm sp} - \frac{1}{2} \equiv
\frac{1}{N}(n_{\rm sp} - \frac{N}{2})$. If $g$ is large enough,
says, $g > g_{\rm sp}$, there is only one saddle point $x_{\rm
sp}$. When $g < g_{\rm sp}$, two saddle points appear, namely
$x_{\rm sp,-} < x_{\rm sp,+}$. $f(x_{\rm sp,+}) > f(x_{\rm sp,-})$
as $p < \frac{1}{2}$ and vice versa. The plot of the saddle points
as a function of $g$ for different $p$ are shown in Fig.
\ref{y_vs_g.eps}. $g_{\rm sp}$ as a function of $p$ is plotted in
the inset.

\vspace{25pt}

The numerator in Eq. (\ref{nmean}) in large $N$ limit can be
evaluated by a similar way. In large $N$ limit, $\langle n \rangle
= n_{\rm sp}$ if $g > g_{\rm sp}$. When $g < g_{\rm sp}$, we have
\begin{eqnarray} \label{n-1st}
\langle n \rangle = \left\{ \begin{tabular}{cc}
  $n_{\rm sp,+}$ & \ \ \ if \ $p<\frac{1}{2}$  \\
  $\frac{N}{2}$ & \ \ \ if \ $p=\frac{1}{2}$ \\
  $n_{\rm sp,-}$ & \ \ \ if \ $p>\frac{1}{2}$
\end{tabular} \right.
\end{eqnarray}

When $p=\frac{1}{2}$, the system undergoes a second-order phase
transition by varying the coupling constant $g$. The order
parameter, $\langle n \rangle$, changes continuously across the
transition. The critical point $g_c$ can be determined by solving
$f''(x_{\rm sp})|_{g\rightarrow g_c^+}=0$, which gives $g_c=-2$.

\begin{figure}[tbh]
\begin{center}
\includegraphics[width=3in]{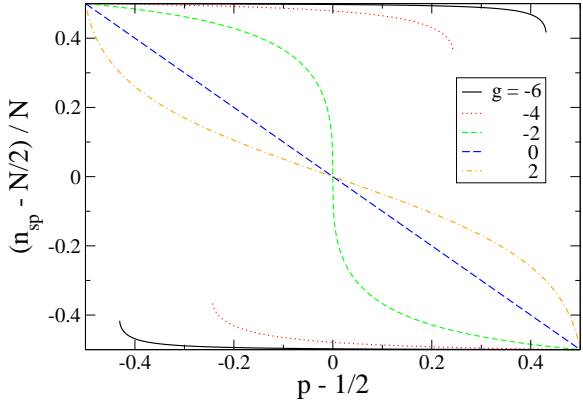}
\end{center}
\vspace{-5pt}
 \caption{The relation between the saddle
point $y_{\rm sp}$ in Eq. (\ref{y}) and $p$. $\frac{1}{N}(n_{\rm
sp} -\frac{N}{2})$ as a function of $p-\frac{1}{2}$ for different
$g$.
 }
 \label{y_vs_p.eps}
 \vspace{25pt}
\end{figure}

If $g < g_c$, there's a first-order phase transition as $p$
varies. The critical point $p_c$ is given by $f(x_{\rm
sp,+})|_{p\rightarrow p_c^-} = f(x_{\rm sp,-})|_{p\rightarrow
p_c^+}$, which gives $p_c=\frac{1}{2}$. As seen from Eq.
(\ref{n-1st}) and Fig. \ref{y_vs_p.eps}, the order parameter,
$\langle n \rangle$, changes discontinuously at $p=p_c$. The
saddle point at $x_{\rm sp,-} ( x_{\rm sp,+} ) $ when $p < p_c ( p
> p_c )$ represents the metastable state. Due to the existence of
the metastable state, the system shows hysteresis. In the section
VI, we provide another means to indicate the existence of
metastability.

\vspace{25pt}

\section{IV. Relaxation to equilibrium}

When the system is not at its equilibrium, it will relax. It is
interesting to know how the relaxation time behaves. Expanding Eq.
(\ref{ns}) with the help of Eq. (\ref{ninfty}) gives
\begin{eqnarray} \label{relax}
\langle n \rangle_s &=& \langle n \rangle + \lambda_{N-1}^s
\sum_{n,k=0}^N n A_{n,N-1}  A_{N-1,k}^{-1} \psi_k(0) \nonumber \\
&& + \lambda_{N-2}^s
\sum_{n,k=0}^N n A_{n,N-2}  A_{N-2,k}^{-1} \psi_k(0) \nonumber \\
&& + \cdots + \lambda_0^s \sum_{n,k=0}^N n A_{n0} A_{0k}^{-1}
\psi_k(0)
\end{eqnarray}
where the eigenvalues are arranged in ascending order, $\lambda_0
< \lambda_1 < \cdots < \lambda_{N-1} < \lambda_N = 1$. If $s$ is
large enough, the last contribution term before reaching the
equilibrium is the $\lambda_{N-1}$ term, which also defines the
relaxation time
\begin{eqnarray} \label{taur}
\tau_{\rm R} \equiv -\frac{1}{\ln (\lambda_{N-1})}
\end{eqnarray}
If $g = 0$, for large $N$, $\tau_{\rm R}=N$. If $g \neq 0$, the
eigenvalues can be found by perturbation (See Appendix D for
details).

For $g > g_{\rm sp}$, let $M_0$ be the transition matrix at $g=0$.
The perturbed transition matrix $M_1 = M - M_0$, and then apply
Eq. (\ref{eigenvaluefirst}), after some algebra, we have the
first-order perturbation correction to the $m$-th eigenvalue
\begin{widetext}
\begin{eqnarray} \label{lambdam}
\lambda_m^{(1)} &=& -\frac{1}{2 N}\sum_{n=1}^N  A^{-1}_{mn} (
q (N-n+1) A_{n-1,m}  + p n  A_{nm} )  \tanh[\frac{g}{2N}(2n-N-1)] \nonumber \\
&&  + \frac{1}{2 N}\sum_{n=0}^{N-1}  A^{-1}_{mn} ( q (N-n) A_{nm}
+ p (n +1)  A_{n+1,m} )  \tanh[\frac{g}{2N}(2n-N+1)]  \nonumber \\
&=& \frac{1}{2 N} \sum_{k=0}^{\frac{N}{2}-1}
\tanh[\frac{g}{2N}(2k+1)] \left( [A^{-1}_{m,\frac{N}{2}-k} -
A^{-1}_{m,\frac{N}{2}-k-1}] [ q (\frac{N}{2}+k+1)
A_{\frac{N}{2}-k-1,m} + p (\frac{N}{2}-k) A_{\frac{N}{2}-k,m} ]
\right. \nonumber \\
&& \left. + [A^{-1}_{m,\frac{N}{2}+k} -
A^{-1}_{m,\frac{N}{2}+k+1}] [ q (\frac{N}{2}-k)
A_{\frac{N}{2}+k,m} + p (\frac{N}{2}+k+1) A_{\frac{N}{2}+k+1,m} ]
\right)
\end{eqnarray}
\end{widetext}

For $m=N$, notice that $A^{-1}_{Nn}=1$ by using Eqs.
(\ref{phinm2}) and (\ref{anminv}), we get $\lambda_N^{(1)}=0$. It
is consistent with the fact that the eigenvalue $\lambda_N = 1$
for the equilibrium state should be unchanged under perturbation.

The next largest eigenvalue is responsible for the relaxation time
to the equilibrium. For $m = N-1$ in Eq. (\ref{lambdam}), and
notice that $A^{-1}_{N-1,n}=qN-n$ and
$A_{n,N-1}=\phi_n(N)(qN-n)/(Npq)$ which can be obtained from Eqs.
(\ref{phinm2}) and (\ref{anminv}), after some algebra, we have
\begin{widetext}
\begin{eqnarray}
\lambda_{N-1}^{(1)} &=& -\frac{1}{2q
N^2}\sum_{k=0}^{\frac{N}{2}-1}
\frac{N!}{(\frac{N}{2}+k)!(\frac{N}{2}-k)!} p^{\frac{N}{2}+k}
q^{\frac{N}{2}-k} (\frac{N}{2}-k) [2k+1+(2q-1)N]
\tanh[\frac{g}{2N}(2k+1)] \nonumber \\
&& -\frac{1}{2p N^2}\sum_{k=0}^{\frac{N}{2}-1}
\frac{N!}{(\frac{N}{2}+k)!(\frac{N}{2}-k)!} q^{\frac{N}{2}+k}
p^{\frac{N}{2}-k} (\frac{N}{2}-k) [2k+1+(2p-1)N]
\tanh[\frac{g}{2N}(2k+1)] \nonumber \\
&\simeq& -\frac{1}{q}\sqrt{\frac{N}{2\pi p q}}
\int_{-\frac{1}{2}(p-q)}^q dx {\rm e}^{-\frac{N}{2pq}x^2} x (q -
x) \tanh[g(x + \frac{1}{2}(p-q))] \nonumber \\
&& -\frac{1}{p}\sqrt{\frac{N}{2\pi p q}}
\int_{-\frac{1}{2}(q-p)}^p dx {\rm e}^{-\frac{N}{2pq}x^2} x (p -
x) \tanh[g(x + \frac{1}{2}(q-p))] \nonumber \\
&=& -\frac{g pq}{N} {\rm sech}^2[\frac{g}{2}(q-p)] + O(N^{-2})
\end{eqnarray}
\end{widetext}
where we only keep the leading order for large $N$ in the
asymptotic expansion. From the definition of the relaxation time
in Eq. (\ref{taur}),
\begin{eqnarray}
\tau_{\rm R} = \frac{2N}{1 + 2 g pq {\rm
sech}^2[\frac{g}{2}(q-p)]}
\end{eqnarray}
In particular, when $p=\frac{1}{2}$,
\begin{eqnarray}
\tau_{\rm R} = \frac{2 N}{1 + \frac{g}{2}}
\end{eqnarray}
Notice that $\tau_R \rightarrow 0$ as $g\rightarrow +\infty$. The
more repulsive interaction, the shorter relaxation time to the
equilibrium.

\begin{figure}[tbh]
\begin{center}
\includegraphics[width=3in]{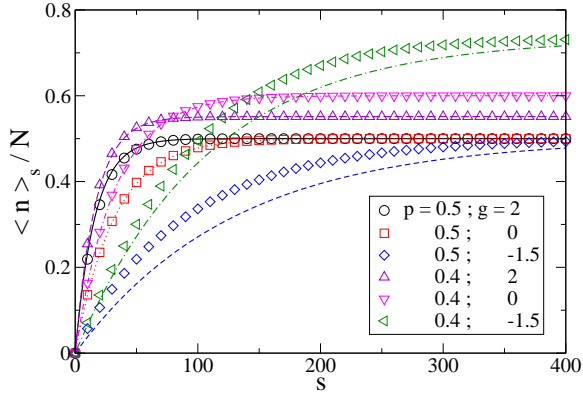}
\end{center}
\vspace{-5pt}
 \caption{The proportion of particle number in the left urn, $\langle n \rangle_s / N$, as a function of time step $s$ for $g > g_{\rm sp}$ at different $p$ and $g$.
  The initial value $n_0$ is chosen to be the most far-from-equilibrium state. Solid lines represent the
 corresponding result by Eq. (\ref{relaxggc}).}
 \label{relaxggc.eps}
 \vspace{25pt}
\end{figure}

By keeping only the first two terms in Eq. (\ref{relax}), and
using the definition of $\tau_R$ from Eq. (\ref{taur}), we have
\begin{eqnarray} \label{relaxggc}
\langle n \rangle_s = \langle n \rangle + (n_0 - \langle n
\rangle) {\rm e}^{-s/\tau_{\rm R}}
\end{eqnarray}
as $s$ is large enough. $n_0$ is the initial value. The above
formula is compared with the numerical result, as shown in Fig.
\ref{relaxggc.eps}. Good agreement at large $s$ is found.

For $g < g_{\rm sp}$, let $M_0$ be the transition matrix at
$g\rightarrow -\infty$. Without loss of generality, suppose $p
\leq \frac{1}{2}$, then the equilibrium eigenstate is labelled by
$m=N$, of eigenvalue $\lambda_N^{(0)}=1$. The eigenstate of the
next largest eigenvalue is labelled by $m=N-1$, of eigenvalue
$\lambda_{N-1}^{(0)} = 1 - \frac{p}{N}$.

By Eq. (\ref{eigenvaluefirst}), and notice that the non-vanishing
$A_{nN}=1$ for $n=N$, $A_{n,N-1}=(-1)^n$ for $n \geq N-1$ from Eq.
(\ref{phinm}), we get
$\lambda_N^{(1)}=(M_1)_{NN}+(M_1)_{N-1,N}=0$, which is again
consistent with $\lambda_N=1$ unchanged under perturbation.

The first-order perturbation correction to the next largest
eigenvalue is
\begin{eqnarray}
\lambda_{N-1}^{(1)} &=& -(M_1)_{N-1,N} + (M_1)_{N-1,N-1} + 2
(M_1)_{N-2,N-1} \nonumber \\ &=& - \frac{q N-p}{N} \frac{1}{{\rm
e}^{|g|(1-\frac{1}{N})} +1} + \frac{q(N-1)}{N} \frac{1}{{\rm
e}^{|g|(1-\frac{3}{N})} +1}  \nonumber \\
&=& \frac{1}{N} \left( \frac{|g| q}{2} {\rm sech}^2(\frac{|g|}{2})
+ \frac{p-q}{{\rm e}^{|g|}+1} \right) + O(N^{-2})
\end{eqnarray}
if only the leading order for large $N$ is kept. The relaxation
time is then
\begin{eqnarray}
\tau_{\rm R} = \frac{N}{p - \frac{|g| q}{2} {\rm
sech}^2(\frac{|g|}{2}) - \frac{p-q}{{\rm e}^{|g|}+1}}
\end{eqnarray}
$g$ is largely negative if $p$ deviates from $\frac{1}{2}$ a lot,
as shown in the inset of Fig. \ref{y_vs_g.eps}. In this case,
$\tau_{\rm R}\simeq \frac{N}{p}$, which is the limit as
$g\rightarrow -\infty$.

Similarly, if $p > \frac{1}{2}$, the relaxation time is
\begin{eqnarray}
\tau_{\rm R} = \frac{N}{q - \frac{|g| p}{2} {\rm
sech}^2(\frac{|g|}{2}) - \frac{q-p}{{\rm e}^{|g|}+1}}
\end{eqnarray}
and $\tau_{\rm R}\simeq \frac{N}{q}$ when $p$ deviates from
$\frac{1}{2}$ a lot.

\begin{figure}[tbh]
\begin{center}
\includegraphics[width=3in]{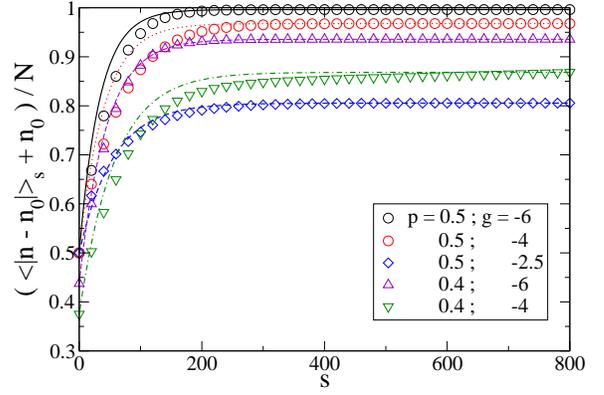}
\end{center}
\vspace{-5pt}
 \caption{The proportion of particle number in the left urn, $(\langle |n - n_0| \rangle_s + n_0)/N$, as a function of time step $s$ for $g < g_{\rm sp}$ at different $p$ and $g$.
  The initial value $n_0$ is chosen to be the most far-from-equilibrium state. Solid lines represent the
 corresponding result by Eq. (\ref{relaxlgc}).}
 \label{relaxlgc.eps}
 \vspace{25pt}
\end{figure}

With the help of the relaxation time, we have
\begin{eqnarray} \label{relaxlgc}
\langle |n - n_0| \rangle_s + n_0 = \langle n \rangle + (n_0 -
\langle n \rangle) {\rm e}^{-s/\tau_{\rm R}}
\end{eqnarray}
as $s$ is large enough. $n_0$ is the initial value. Here we use
$\langle |n - n_0| \rangle_s + n_0$ instead of $\langle n
\rangle_s$ in order to avoid the interference from the metastable
state. The above formula is compared with the numerical result, as
shown in Fig. \ref{relaxlgc.eps}. Again both analytical and
numerical results match well at large $s$.

\vspace{25pt}

\vspace{25pt}

\section{V. Poincare cycle}

In this section, we are going to discuss the scaling behavior of
the Poincare cycle with respect to the particle number $N$ and the
tuning parameters ($g$ or $p$) across the (first-order and
second-order) phase transition.

The Poincare cycle of the state $|n\rangle$, denoted by $\tau_{\rm
P}(n)$, is defined as the mean time from the state $|n\rangle$ to
its original state at its first time, $\tau(n\rightarrow n)$,
which is (see Appendix C for the proof)
\begin{eqnarray}
\tau_{\rm P}(n) = \frac{\sum_{k=0}^N \phi_k(N)}{\phi_n(N)}
\end{eqnarray}
If $p=\frac{1}{2}$, by Eqs. (\ref{phinN}) and (\ref{sumlargeN}),
and also notice that $f''(x_{\rm sp}) = -2(g - g_c)$ with $g_c =
-2$, it is straightforward to have the Poincare cycle of the
equilibrium
\begin{eqnarray}
\tau_{\rm P}^{\rm eq} = \tau_{\rm P}(n_{\rm sp}) = \sqrt{\pi} (g -
g_c)^{-\frac{1}{2}} N^\frac{1}{2}
\end{eqnarray}

If $g < g_c$, two saddle points $n_{{\rm sp},\pm}$ emerge. When $g
\lesssim g_c$, $n_{{\rm sp},+}\gtrsim\frac{N}{2}$ and $n_{{\rm
 sp},-}\lesssim\frac{N}{2}$. $f(x_{{\rm sp},+})=f(x_{{\rm sp},-})$. One can solve
for $x_{{\rm sp},\pm}\simeq \frac{1}{2}\pm \sqrt{\frac{3}{8}}(g_c
- g)^\frac{1}{2}$. Notice that $f''(x_{{\rm sp},\pm}) = -4(g_c -
g)$, the Poincare cycle of the equilibrium
\begin{eqnarray}
\tau_{\rm P}^{\rm eq} = \sqrt{2\pi} (g_c - g)^{-\frac{1}{2}}
N^\frac{1}{2}
\end{eqnarray}
When $g \ll g_c$, $n_{{\rm sp},+}\lesssim N$ and $n_{{\rm sp},-}
\gtrsim 0$, then $x_{{\rm sp},+}\simeq 1 - {\rm e}^g$, and
$x_{{\rm sp},-} \simeq {\rm e}^g$. $f''(x_{{\rm sp},\pm})\simeq -
{\rm e}^{-g}$, then
\begin{eqnarray}
\tau_{\rm P}^{\rm eq} = 2 \sqrt{2\pi} {\rm e}^{-\frac{|g|}{2}}
N^\frac{1}{2}
\end{eqnarray}

The Poincare cycle of the equilibrium state $\tau_{\rm P}^{\rm
eq}$ has always $\sqrt{N}$ dependence. It becomes divergent at the
transition point $g = g_c$.  From Eq. (\ref{sumlargeN}), it is
seen that the divergence comes from the vanishing $|f''(x_{\rm
sp})|_{g\rightarrow g_c}$ in the denominator, which implies that
it is universal in second-order phase transition. However, note
that Eq. (\ref{sumlargeN}) is obtained in large $N$ limit. For
large but finite $N$, one should see the divergent-like scaling
behavior instead of real divergence.

Next we investigate the scaling behavior of the Poincare cycle of
the most far-from-equilibrium state $\tau_{\rm P}^{\rm feq}$, in
which it is defined as the longest Poincare cycle.

When $p=\frac{1}{2}$, and $g > g_c$, then $n_{\rm sp} =
\frac{N}{2}$, the most far-from-equilibrium state is at $n = N$
(or $n = 0$), then
\begin{eqnarray}
\tau_{\rm P}^{\rm feq} = \tau_{\rm P}(N) =\sqrt{2} (g -
g_c)^{-\frac{1}{2}} \exp [ N( \ln 2 + \frac{g}{4} ) ]
\end{eqnarray}
If $g \lesssim g_c$,
\begin{eqnarray}
\tau_{\rm P}^{\rm feq} = \tau_{\rm P}(N) = 2 (g_c -
g)^{-\frac{1}{2}} \exp [ N( \ln 2 + \frac{g}{4} ) ]
\end{eqnarray}
If $g \ll g_c$,
\begin{eqnarray}
\tau_{\rm P}^{\rm feq} = \tau_{\rm P}(\frac{N}{2}) = \sqrt{2 \pi}
N^\frac{1}{2} \exp [ N( \frac{|g|}{4} - \ln 2 ) ]
\end{eqnarray}

The Poincare cycle of the most far-from-equilibrium $\tau_{\rm
P}^{\rm feq}$ has the exponential form
 ${\rm e}^{\alpha N}$. It also becomes ``divergent" (see the argument above) at $g=g_c$,
but the scaling exponent $\alpha$ is finite and continuous across
the transition point.

In the following, we are going to investigate the behavior of the
Poincare cycle across the first-order phase transition. Suppose $g
\ll g_{\rm sp}$, there are two saddle points. When $p < p_c
=\frac{1}{2}$, $x_{\rm sp,+} \simeq 1 - (1 + \ln\frac{p}{q}) {\rm
e}^g$ and $x_{\rm sp,-} \simeq (1 - \ln\frac{p}{q}) {\rm e}^g$,
the Poincare cycle of the equilibrium state
\begin{eqnarray}
\tau_{\rm P}^{\rm eq} = \tau_{\rm P}(n_{\rm sp,+})
 = \sqrt{2\pi} {\rm e}^{-\frac{|g|}{2}} N^\frac{1}{2}
 \left(1 + \ln\left(\frac{p}{q}\right)\right)^\frac{1}{2}
\end{eqnarray}

When $p > p_c$, $x_{\rm sp,+} \simeq 1 - (1 + \ln\frac{q}{p}) {\rm
e}^g$, $x_{\rm sp,-} \simeq (1 - \ln\frac{q}{p}) {\rm e}^g$, then
\begin{eqnarray}
\tau_{\rm P}^{\rm eq} = \tau_{\rm P}(n_{\rm sp,-})
 = \sqrt{2\pi} {\rm e}^{-\frac{|g|}{2}} N^\frac{1}{2}
 \left(1 + \ln\left(\frac{q}{p}\right)\right)^\frac{1}{2}
\end{eqnarray}

It is interesting to notice that $\tau_{\rm P}^{\rm eq} \simeq
\sqrt{2\pi} |g|^{-\frac{1}{2}} N^\frac{1}{2}$ if $p\neq p_c$. At
the transition point $p=p_c$, $\tau_{\rm P}^{\rm eq} = 2
\sqrt{2\pi} |g|^{-\frac{1}{2}} N^\frac{1}{2}$. The Poincare cycle
of the equilibrium state is finite and continuous during
first-order transition.

The most far-from-equilibrium state is at $n = N(\frac{1}{2}
-\ln(\frac{p}{q}) {\rm e}^g)$, then
\begin{eqnarray}
\tau_{\rm P}^{\rm feq} = && \sqrt{\frac{\pi}{2}}  N^\frac{1}{2}
\exp [ N( \frac{|g|}{4} -
\ln 2 ) ] \nonumber \\
&& \times \left( \left(\frac{p}{q} \right)^\frac{N}{2} +
\left(\frac{q}{p} \right)^\frac{N}{2} \right)
\end{eqnarray}

When $p$ is around the transition point $p_c$, in large $N$ limit,
$\tau_{\rm P}^{\rm feq} = \sqrt{\frac{\pi}{2}} N^\frac{1}{2} \exp
[ N( \frac{|g|}{4} - \ln 2 ) ] $. At exactly $p = p_c$, $\tau_{\rm
P}^{\rm feq}$ is double its value.

The Poincare cycle of the most far-from-equilibrium has still the
exponential form ${\rm e}^{\alpha N}$ dependence, with a
continuous exponent $\alpha$ across the first-order phase
transition.

In summary, the Poincare cycles $\tau_{\rm P}^{\rm eq}$ and
$\tau_{\rm P}^{\rm feq}$ have the $\sqrt{N}$ and ${\rm e}^{\alpha
N}$ dependence, respectively. During second-order phase
transition, both $\tau_{\rm P}^{\rm eq}$ and $\tau_{\rm P}^{\rm
feq}$ behave divergent-like at the transition point. At
first-order phase transition, the Poincare cycles are finite and
continuous. Such a behavior of the Poincare cycle could be served
as an indication of the nature of the phase transition


\section{VI. Duration Time}

When $g \ll g_{\rm sp}$, the system will stay at the states
$|0\rangle$ and $|N \rangle$. Suppose the system transits from
$|N\rangle$ to $|0\rangle$, it should meet $|\frac{N}{2}\rangle$
during the evolution because $n$ changes continuously (Here the
continuity of $n$ means $n$ changes its value at most $\pm 1$ at
each step).

Define $\tau_{\rm D}(n,1)$ as the mean time for the system to
evolve from $|\frac{N}{2}\rangle$ to $|n\rangle$ at its first
time, and then back $|\frac{N}{2}\rangle$ at its first time. When
$n=N$,
\begin{eqnarray} \label{taud1}
\tau_{\rm D}(N,1) \equiv \sum_{s_1=1}^\infty \sum_{s_2=1}^\infty
(s_1+s_2) (\frac{N}{2}|p(s_2)|N) (N|p(s_1)|\frac{N}{2}) \nonumber
\\
\end{eqnarray}
where the notation $(m|p(s)|n)$ represents the probability that
the state $|m\rangle$ becomes $|n\rangle$ at its first time after
$s$ steps. With the help of Eqs. (\ref{gnm}-\ref{taup}), Eq.
(\ref{taud1}) becomes
\begin{eqnarray}
\tau_{\rm D}(N,1) &=& \tau(\frac{N}{2}\rightarrow N)
g_{\frac{N}{2},N}(1) + \tau(N\rightarrow \frac{N}{2})
g_{N,\frac{N}{2}}(1) \nonumber \\
&=& \tau_{\rm P}(N) + \tau_{\rm P}(\frac{N}{2})
\end{eqnarray}
Since $\tau_{\rm P}(\frac{N}{2}) \gg \tau_{\rm P}(N)$ for $g \ll
g_{\rm sp}$, $\tau_{\rm D}(N,1) = \tau_{\rm P}(\frac{N}{2})$. By
similar argument, $\tau_{\rm D}(0,1) = \tau_{\rm P}(\frac{N}{2})$.

The above transition ( $|\frac{N}{2}\rangle \rightarrow |n\rangle
\rightarrow |\frac{N}{2}\rangle$ ) may occur $k$ times
consecutively. Define $\tau_{\rm D}(n,k)$ as its mean time, then
\begin{widetext}
\begin{eqnarray}
\tau_{\rm D}(n,k) &\equiv& \sum_{s_1,\ldots,s_{2k} = 1}^\infty
(s_1 + s_2 + \ldots + s_{2k}) (\frac{N}{2}|p(s_{2k}|n)
(n|p(s_{2k-1})|\frac{N}{2}) \cdots (\frac{N}{2}|p(s_2)|n)
(n|p(s_1)|\frac{N}{2}) \nonumber \\
&=& k \sum_{s=1}^\infty s (n|p(s)|\frac{N}{2}) + k
\sum_{s=1}^\infty s (\frac{N}{2}|p(s)|n) \nonumber \\
&=& k \left(\tau_{\rm P}(n) + \tau_{\rm P}(\frac{N}{2})\right)
\end{eqnarray}
\end{widetext}
Hence $\tau_{\rm D}(N,k) = \tau_{\rm D}(0,k) = k \tau_{\rm
P}(\frac{N}{2})$.

The duration time at state $|N\rangle$, $\tau_D(N)$, defined as
the total time at which the system stays at $|N\rangle$ before
transits to $|0\rangle$,
\begin{eqnarray}
\tau_{\rm D}(N) &\equiv& \sum_{k=1}^\infty \left(\frac{\tau_{\rm
P}(N)^{-1}}{\tau_{\rm P}(N)^{-1} + \tau_{\rm P}(0)^{-1}}\right)^k
\tau_{\rm D}(N,k) \nonumber \\
&=& \left(\tau_{\rm P}(N) + \tau_{\rm P}(\frac{N}{2})\right)
\sum_{k=1}^\infty k
\left(\frac{p^N}{p^N+q^N}\right)^k  \nonumber \\
&=& \left(\frac{q}{p}\right)^N \left[\left(\frac{q}{p}\right)^N +
1\right] \left(\tau_{\rm
P}(N) + \tau_{\rm P}(\frac{N}{2})\right) \nonumber \\
\end{eqnarray}
The asymptotic form at large $N$ limit becomes
\begin{eqnarray} \label{taudN}
\tau_{\rm D}(N) = \left\{ \begin{tabular}{cc}
  $(\frac{q}{p})^{2N} \tau_{\rm P}(\frac{N}{2})$ & \ \ \ if \ $p < \frac{1}{2}$ \\
  $2 \tau_{\rm P}(\frac{N}{2})$ & \ \ \ if \ $p = \frac{1}{2}$ \\
  $(\frac{q}{p})^N \tau_{\rm P}(\frac{N}{2})$ & \ \ \ if \ $p > \frac{1}{2}$ \\
\end{tabular} \right.
\end{eqnarray}

Similarly, the duration time at state $|0\rangle$ is
\begin{eqnarray}
\tau_{\rm D}(0) = \left(\frac{p}{q}\right)^N
\left[\left(\frac{p}{q}\right)^N + 1\right] \left(\tau_{\rm
P}(0) + \tau_{\rm P}(\frac{N}{2})\right) \nonumber \\
\end{eqnarray}
and its asymptotic form
\begin{eqnarray}
\tau_{\rm D}(0) = \left\{ \begin{tabular}{cc}
  $(\frac{p}{q})^N \tau_{\rm P}(\frac{N}{2})$ & \ \ \ if \ $p < \frac{1}{2}$ \\
  $2 \tau_{\rm P}(\frac{N}{2})$ & \ \ \ if \ $p = \frac{1}{2}$ \\
  $(\frac{p}{q})^{2N} \tau_{\rm P}(\frac{N}{2})$ & \ \ \ if \ $p > \frac{1}{2}$ \\
\end{tabular} \right.
\end{eqnarray}

There is a first-order phase transition as $p$ varies. As $p <
p_c$, $\tau_{\rm D}(N) > \tau_{\rm D}(0) > 0$. It means the state
$|N\rangle$ is preferable but $|0\rangle$ still survives. Upon
increasing $p$, the ratio of the duration time of two states,
$\tau_{\rm D}(N)/\tau_{\rm D}(0)$, decreases. At $p=p_c$,
$\tau_{\rm D}(N) = \tau_{\rm D}(0)$. Further increasing $p > p_c$,
$\tau_{\rm D}(0) > \tau_{\rm D}(N) > 0$. Such a behavior indicates
a strong evidence of metastability during first-order phase
transition.


\section{VII. Discussion}


The order-of-magnitude determination of the Poincare cycle of the
most far-from-equilibrium state was originally used to resolve the
recurrence paradox. In macroscopic world, it is far beyond the
time scale we can observe. If $N$ is not large enough, in
principle, the measurement of the Poincare cycle should be
experimentally accessible. For example, in colloidal system, one
can easily prepare the system of small particle number $N$. The
interaction between the colloidal particles ($g$ in our model) is
also well controlled \cite{yethiraj}. The probability of directed
transport ($p$ in our model) can be tuned by applying the electric
field along the direction from the left to the right urn, and the
particles are slightly charged.


\section{acknowledgement}
One of the authors (C.H.C.) thanks Pik-Yin Lai and Chi-Ning Chen
for their helpful discussion. The work was supported by the
Ministry of Science and Technology of the Republic of China.

\appendix*
\section{Appendix A: Eigenproblem for $g=0$}

The standard way to solve the eigenproblem is the method of
generating functions \cite{feller}. For $g=0$, it was already
known \cite{siegert,klein,lee}. In the following, we briefly
outline the solution.

For $g=0$, Eq. (\ref{eigensystem}) is reduced to
\begin{widetext}
\begin{eqnarray}
&& \frac{N-n+1}{2 N} q \phi_{n-1} + \frac{n+1}{2 N} p \phi_{n+1} +
\left( 1 - \frac{n}{2 N} p - \frac{N-n}{2 N} q \right)\phi_n =
\lambda \phi_n
\end{eqnarray}
\end{widetext}
Let $f(z)\equiv\sum_{n=0}^N \phi_n z^n=\sum_{n=-\infty}^\infty
\phi_n z^n$, if we extend $\phi_n\equiv 0$ for $n<0$ and $n>N$,
then
\begin{eqnarray}
\frac{1}{N}\frac{d f}{d z} (p + q z)(1-z) = (2\lambda -1 - (p+qz))
f
\end{eqnarray}
The solution is
\begin{eqnarray}
f(z) = (p+ q z)^{N (2\lambda -1)} (1-z)^{2 N(1-\lambda)}
\label{fz}
\end{eqnarray}
up to an arbitrary proportional constant. Since $f(z)$ is a
polynomial in $z$ by definition, $N (2\lambda -1)$ and $2
N(1-\lambda)$ have to be non-negative integers. Hence we get
\begin{eqnarray}  \label{lmg0}
\lambda_m = \frac{1}{2}+\frac{m}{2 N}
\end{eqnarray}
where $m=0,1,2, \ldots , N$ are the numbers to label the
eigenvalues. The corresponding eigenvectors of the component
$\phi_n(m)$ could be obtained by comparing the $z^n$ coefficient
of $f(z)$ in Eq. (\ref{fz}) with its definition, we have
\begin{eqnarray} \label{phinm2}
\phi_n(m) = \sum_{k+l=n} \left( \begin{array}{c}
  m \\
  k \\
\end{array}
\right)
\left(
\begin{array}{c}
  N-m \\
  l \\
\end{array}
\right) (-1)^l p^{m-k}q^k \nonumber \\
\end{eqnarray}
In particular, for $\lambda_N = 1$, its corresponding eigenvector
\begin{eqnarray}
\phi_n(N) = \frac{N!}{n! (N-n)!} p^{N-n} q^n
\end{eqnarray}

Now $A_{nm} = \phi_n(m)$, its inverse $A_{nm}^{-1}$ is defined as
\begin{eqnarray}
\sum_n A_{ln} A_{nm}^{-1} = \delta_{lm}
\end{eqnarray}
Multiply $z^l$, sum over $l$, and make use of
Eq.(\ref{fz}-\ref{lmg0}), we get
\begin{eqnarray}
\sum_n (p + q z)^n (1-z)^{N-n} A_{nm}^{-1} = z^m
\end{eqnarray}
By change of variable $t = -\frac{p+q z}{1-z}$, we have
\begin{eqnarray}
\sum_n A_{nm}^{-1} (-1)^{m+n} t^n = q^{N-m} f_m(\frac{t}{q})
\end{eqnarray}
which gives
\begin{eqnarray} \label{anminv}
A_{nm}^{-1} = (-1)^{m+n} q^{N-m-n} \phi_n(m)
\end{eqnarray}

\section{Appendix B: Eigenproblem for $g \rightarrow -\infty$}

As $g \rightarrow -\infty$, Eq. (\ref{eigensystem}) is reduced to
\begin{widetext}
\begin{eqnarray}
\frac{N-n+1}{N} q \Theta(n-\frac{N+1}{2}) \phi_{n-1} +
\frac{n+1}{N} p \Theta(\frac{N-1}{2}-n) \phi_{n+1} \nonumber \\ +
\left( 1 - \frac{n}{N} p \Theta(\frac{N+1}{2}-n) -\frac{N-n}{N} q
\Theta(n-\frac{N-1}{2}) \right) \phi_n = \lambda \phi_n
\label{gminfty}
\end{eqnarray}
\end{widetext}
where $\Theta(x)$ is the step function. When $n=\frac{N}{2}$, it
becomes $\frac{1}{2} \phi_{\frac{N}{2}}= \lambda
\phi_{\frac{N}{2}}$. Hence $\lambda_\frac{N}{2} = \frac{1}{2}$,
and the corresponding eigenvector is $\phi(\frac{N}{2}) = (0,
\ldots, 1, \ldots, 0)^T$ with the only non-vanishing component
$\phi_{\frac{N}{2}}(\frac{N}{2})=1$ (Here we label this eigenstate
by $\frac{N}{2}$ ). The matrix $M$ is in block diagonal form. We
first search for the eigenstates such that $\phi_n=0$ for $n\geq
\frac{N}{2}$, and further assume that $\phi_n \equiv 0$ for $n<0$.
Let $f(z)\equiv\sum_{n=0}^N \phi_n z^n=\sum_{n=-\infty}^\infty
\phi_n z^n$, then
\begin{eqnarray}
\frac{p}{N}\frac{df}{dz}(1-z) = (\lambda-1)f
\end{eqnarray}
The solution is
\begin{eqnarray} \label{fzgm}
f(z) = (1-z)^{N(1-\lambda)/p}
\end{eqnarray}
up to an arbitrary proportional constant. Since $f(z)$ is a
polynomial of degree $\frac{N}{2}-1$ in $z$ by definition, $N
(1-\lambda)/p$ have to be non-negative integers less than or equal
to $\frac{N}{2}-1$. Hence we get
\begin{eqnarray} \label{lmgm}
\lambda_m = 1- p\frac{N-m}{N}
\end{eqnarray}
where $m=\frac{N}{2}+1, \ldots, N-1, N$ are the numbers to label
the eigenvalues. The non-vanishing components of the corresponding
eigenvectors are
\begin{eqnarray}
\phi_n(m) = (-1)^n \left(\begin{array}{c}
  N-m \\
  N-n \\
\end{array} \right)  \label{phinm}
\end{eqnarray}
which are the $z^{N-n}$ coefficient of $f_m(z) = (1- z)^{N-m}$
with $\frac{N}{2}+1 \leq m \leq n \leq N$.

By making the transformation from $n$ to $N-n$ and $p$ to $1-p$ in
Eq.(\ref{gminfty}), we get another set of eigenstates such that
\begin{eqnarray}
\phi_n(m) = \phi_{N-n}(N-m)
\end{eqnarray}
where $m = 0, 1, \ldots, \frac{N}{2}-1$. With the help of
Eq.(\ref{lmgm}-\ref{phinm}), the non-vanishing components of the
eigenvectors are
\begin{eqnarray} \label{phinm3}
\phi_n(m) = (-1)^n \left(\begin{array}{c}
  m \\
  n \\
\end{array} \right)
\end{eqnarray}
where $0 \leq n \leq m \leq \frac{N}{2}-1$, $f_m(z)=(1-z)^m$, and
the corresponding eigenvalues are
\begin{eqnarray} \label{lmgm3}
\lambda_m = 1- q\frac{m}{N}
\end{eqnarray}

Now the matrix $A_{nm} = \phi_n(m)$ is block diagonal with three
blocks, $\{ A_{nm} \}_{0\leq n,m \leq \frac{N}{2}-1}$,
$A_{\frac{N}{2},\frac{N}{2}}$, and $\{ A_{nm}
\}_{\frac{N}{2}+1\leq n,m \leq N}$. We first restrict the upper
block, its inverse $A_{nm}^{-1}$ is defined as
\begin{eqnarray}
\sum_{n=0}^{\frac{N}{2}-1} A_{ln} A_{nm}^{-1} = \delta_{lm}
\end{eqnarray}
Similar to the treatment for the case that $g=0$, multiply $z^l$,
sum over $l$, make use of Eq. (\ref{phinm3}-\ref{lmgm3}), and then
make the change of variable $t = 1-z$, we arrive at
\begin{eqnarray}
\sum_{n=0}^{\frac{N}{2}-1} A_{nm}^{-1} t^n = f_m(t)
\end{eqnarray}
which gives
\begin{eqnarray} \label{anm}
A_{nm}^{-1} = \phi_n(m)
\end{eqnarray}
where $0\leq n,m \leq \frac{N}{2}-1$. Eq. (\ref{anm}) also holds
for $0 \leq n,m \leq N$. By the symmetry argument as above, the
transformation $n \rightarrow N-n$, $p \rightarrow 1-p$ leaves Eq.
(\ref{gminfty}) unchanged, Eq. (\ref{anm}) should hold for
$\frac{N}{2}+1 \leq n,m \leq N$. It's also straightforward to
check $A_{\frac{N}{2},\frac{N}{2}}^{-1} =
A_{\frac{N}{2},\frac{N}{2}} = \phi_\frac{N}{2}(\frac{N}{2}) = 1$,
which is Eq. (\ref{anm}) with $n=m=\frac{N}{2}$.

\section{Appendix C: Mean Time from state to state}

Denote $(n|p(s)|m)$ as the probability that the state $|m\rangle$
becomes the state $|n\rangle$ at its first time after $s$ steps.
It's relation with the probability $\langle n|p(s)|m\rangle$ is
\begin{eqnarray} \label{npm}
\langle n|p(s)|m\rangle = (n|p(s)|m) + \sum_{k=1}^{s-1} \langle
n|p(s-k)|n\rangle (n|p(k)|m) \nonumber \\
\end{eqnarray}

Define two generating functions,
\begin{eqnarray}
h_{mn}(z) &\equiv& \sum_{s=1}^\infty \langle n|p(s)|m \rangle z^s
\nonumber \\
&=& \sum_{s=1}^\infty (M^s)_{nm} z^s \nonumber \\
&=& \sum_{s=1}^\infty \sum_{k=0}^N A_{nk} \lambda_k^s A_{km}^{-1}
z^s  \nonumber \\
&=& \sum_{k=0}^N A_{nk} A_{km}^{-1} \frac{\lambda_k z}{1-\lambda_k
z}
\end{eqnarray}
and
\begin{eqnarray}
g_{nm}(z) &\equiv& \sum_{s=1}^\infty (n|p(s)|m) z^s
\end{eqnarray}

We can deduce the relation between these two generating functions
from Eq. (\ref{npm}).
\begin{eqnarray}
h_{mn}(z) = g_{nm}(z) + h_{nn}(z) g_{nm}(z)
\end{eqnarray}
or equivalently,
\begin{eqnarray}
g_{mn}(z) = \frac{h_{nm}(z)}{h_{nn}(z)+1}
\end{eqnarray}

The probability normalization
\begin{widetext}
\begin{eqnarray} \label{gnm}
\sum_{s=1}^\infty (n|p(s)|m) = g_{nm}(1) = \lim_{z\rightarrow 1^-}
\frac{h_{nm}(z)}{h_{nn}(z)+1} = \frac{A_{nN} A_{Nm}^{-1}}{A_{nN}
A_{Nn}^{-1}} = 1
\end{eqnarray}
\end{widetext}
Here we use the fact that $A_{Nk}^{-1}$ is independent of $k$, and
we label $\lambda_N=1$.

The mean time from the state $|m\rangle$ to the state $|n\rangle$
at its first time is defined as
\begin{widetext}
\begin{eqnarray} \label{taumn}
\tau(m \rightarrow n) \equiv \sum_{s=1}^\infty s(n|p(s)|m) =
\left.\frac{d g_{nm}}{dz}\right|_{z=1} = \frac{A_{nN}
A_{Nm}^{-1}}{(A_{nN} A_{Nn}^{-1})^2} = \frac{\sum_{k=0}^N
\phi_k(N)}{\phi_n(N)}
\end{eqnarray}
\end{widetext}
Note that the mean time is independent of the initial state
$|m\rangle$. The Poincare cycle $\tau_{\rm P}(n)$, defined as
$\tau(n\rightarrow n)$, also shares the same result,
\begin{eqnarray} \label{taup}
\tau_{\rm P}(n) = \frac{\sum_{k=0}^N \phi_k(N)}{\phi_n(N)}
\end{eqnarray}

\section{Appendix D: Perturbation Theory}

We want to solve the eigenproblem
\begin{eqnarray} \label{eigenfull}
M \phi(m) = \lambda_m \phi(m)
\end{eqnarray}
Suppose the eigenproblem $M_0 \phi^{(0)}(m) = \lambda_m^{(0)}
\phi^{(0)}(m)$ are solved. Let the matrix
$A_{nm}=\phi^{(0)}_n(m)$, $\psi^{(0)}(m)=A^{-1} \phi^{(0)}(m)$,
then
\begin{eqnarray}
\Lambda_0 \psi^{(0)}(m) = \lambda_m^{(0)} \psi^{(0)}(m)
\end{eqnarray}
where $(\Lambda_0)_{nm}=\lambda_m^{(0)}\delta_{nm}$ and
$\psi_n^{(0)}(m)=\delta_{nm}$. It is obvious to see the
orthnormality relation $\psi^{(0) T}(n) \psi^{(0)}(m) =
\delta_{nm}$.

Write $M = M_0 + M_1$ and $\phi(m) = \phi^{(0)}(m)+\phi^{(1)}(m)$.
Keep Eq. (\ref{eigenfull}) up to the first order, we have
\begin{eqnarray} \label{eigenfirst}
( \Lambda_1 - \lambda_m^{(1)} ) \psi^{(0)}(m) = ( \lambda_m^{(0)}
- \Lambda_0 ) \psi^{(1)}(m)
\end{eqnarray}
where $\Lambda_1 = A^{-1}M_1 A$ and $\psi^{(1)}(m)=A^{-1}
\phi^{(1)}(m)$. Multiply both sides of Eq. (\ref{eigenfirst}) by
$\psi^{(0) T}(m)$, then we get the first order correction of
eigenvalue
\begin{eqnarray} \label{eigenvaluefirst}
\lambda_m^{(1)} = (\Lambda_1)_{mm} = (A^{-1}M_1 A)_{mm}
\end{eqnarray}

\end{document}